\documentclass[conference]{IEEEtran}
\IEEEoverridecommandlockouts
\usepackage{cite}
\usepackage{amsmath,amssymb,amsfonts}
\usepackage{graphicx}
\usepackage{textcomp}
\usepackage{xcolor}
\usepackage{booktabs}
\usepackage{url}
\usepackage{enumitem}
\usepackage{hyperref}
\def\BibTeX{{\rm B\kern-.05em{\sc i\kern-.025em b}\kern-.08em
    T\kern-.1667em\lower.7ex\hbox{E}\kern-.125emX}}
\makeatletter
\def\ps@IEEEtitlepagestyle{%
  \def\@oddhead{}%
  \def\@evenhead{}%
  \def\@oddfoot{\hfil\thepage\hfil}%
  \def\@evenfoot{\hfil\thepage\hfil}}
\makeatother

\begin{document}

\title{Demystifying the Mythos or Disrupting Bugonomics? \\
{\footnotesize From Zero-Day Asymmetry to Defender Remediation Throughput}}






\author{
\begin{tabular}{c@{\hspace{1.2cm}}c@{\hspace{1.2cm}}c}
\begin{tabular}{c}
Alfredo Pesoli\\
\textit{Bynario}\\
{\small alfredo@bynar.io}
\end{tabular}
&
\begin{tabular}{c}
Herman Errico\\
\textit{Vanta}\\
{\small herman.errico@vanta.com}
\end{tabular}
&
\begin{tabular}{c}
Lorenzo Cavallaro\\
\textit{University College London / Bynario}\\
{\small \{l.cavallaro,lorenzo\}@\{ucl.ac.uk,bynar.io\}}
\end{tabular}
\end{tabular}
}

\maketitle
\pagestyle{plain}
\thispagestyle{plain}

\begin{abstract}
Recent demonstrations of large language models producing candidate and confirmed vulnerabilities in production software have renewed the narrative that AI will reshape offensive and defensive security. Headlines emphasize capability; they rarely interrogate costs and incentives. This paper examines LLM-driven vulnerability discovery through a \emph{bugonomics} lens: the operational economics of producing, proving, prioritizing, and fixing security-relevant defects. Historically, the most visible high-end bugonomics was offense-priced because production-grade zero-days and exploit chains were expensive specialist outputs for governments, brokers, and offensive vendors. Defender-side bugonomics already existed in Project Zero-style research, vulnerability reward programs, and vendor remediation work; LLM-assisted systems change its scale and distribution. They make candidate generation, code comprehension, harness construction, test-case generation, proof-of-impact drafting, and report preparation cheaper for defenders operating at codebase scale. Exploits and proofs of concept remain important, but in defender workflows they primarily prove impact, guide prioritization, and justify remediation. The resulting bottleneck is not only finding more bugs; it is absorbing, validating, triaging, patching, and shipping a larger stream of both useful findings and low-quality reports. Using public data from Anthropic's Mythos Preview and Mozilla Firefox collaborations, together with public exploit-market price anchors, vulnerability reward programs, and incident baselines, we argue that the near-term shift is not simply ``more zero-days.'' It is a move from zero-day asymmetry toward broader defender remediation throughput: low-signal candidates become cheaper, evidence-rich remediation packages become more important, and scarce capacity shifts toward maintainer review and release work. The effect is especially acute in open source, where LLM-assisted discovery can increase report volume at machine speed while maintainer-side discovery, validation, triage, funding, and release capacity do not scale automatically with it.
\end{abstract}


\section{Introduction}
In 2026, public demonstrations of frontier language models discovering vulnerabilities in mature open-source projects intensified an already energetic debate about AI and cybersecurity. Anthropic's Mythos Preview reported thousands of previously unknown candidate vulnerabilities across major operating systems and browsers, including long-lived bugs in OpenBSD and FFmpeg and autonomous exploit-development demonstrations \cite{anthropic-mythos}. Anthropic also reported that Claude Opus 4.6 discovered 22 Firefox vulnerabilities over two weeks, 14 of which Mozilla assigned high severity \cite{anthropic-mozilla}. Mozilla later reported 271 Mythos-identified bugs for Firefox 150, including 180 sec-high issues, and 423 total Firefox security bugs fixed across sources in April 2026 \cite{mozilla-hacks-firefox150}. These announcements are usually framed as capability milestones: a stronger model found more bugs, wrote more proofs of concept, or produced more candidate patches.

That framing is incomplete because the security question is economic as much as technical. Bugs, vulnerabilities, proofs of concept, exploit chains, and accepted fixes are not the same product. They have different buyers, prices, failure modes, and strategic value. For decades, high-end zero-days and production exploit chains were scarce specialist goods. Public reporting on exploit brokers and acquisition programs places modern mobile, browser, messenger, and full-chain exploit payouts in the millions of dollars, with prices rising as platforms become harder to compromise \cite{techcrunch-zero-day-prices}. RAND's study of zero-day vulnerabilities likewise treats zero-days and their exploits as durable assets useful to criminals, militaries, governments, defenders, and academics, not as ordinary software defects \cite{rand-zero-days}. Historically, this made the public high-end pricing narrative look offense-centered: the most visible prices were paid for access, exclusivity, stealth, reliability, and operational timing.

\emph{Bugonomics}, as we use the term, is therefore not merely the accounting problem of estimating how many dollars an LLM call costs. It is the study of how vulnerability value changes when the cost of producing different security artifacts changes.
We use this term as an operational security lens, not as a full market-equilibrium theory of vulnerability markets. The unit of analysis is the cost and incentive structure around artifacts defenders can act on: candidate reports, validated findings, proofs of impact, remediation packages, and accepted fixes.
Before LLMs, the most visible high-end prices were offensive because offensive buyers could directly monetize scarce exploit capability, while defender-side economics were concentrated in well-funded vendors, Project Zero-style research, mature VRPs, and internal remediation programs. LLM-assisted systems broaden the defender side of the market by reducing the cost of activities defenders can actually use: understanding code, generating candidate findings, producing reproducers, creating proofs of impact, drafting fixes, and preparing reports that maintainers can act on. A proof of concept is not only an offensive artifact in this framing; for defenders, it is evidence that a bug matters.

This distinction changes the interpretation of Mythos-class results. LLM-assisted systems can lower the cost of reading unfamiliar code, generating hypotheses, writing harnesses, producing minimal test cases, drafting reports, and exploring many targets in parallel. That is economically important because it increases defender leverage. It does not imply that every generated candidate has offensive value, or that the price of a production exploit chain collapses to model-inference cost. The likely near-term effect is a defender-side bottleneck shift: low-signal candidate reports become cheap, evidence-rich remediation packages become more valuable, and scarce effort moves toward validation, prioritization, patch review, and release integration.

We take a deliberately non-adversarial position in the current debate. Frontier models matter: the public results are technically impressive and should be taken seriously. Open-weight models also matter: defenders cannot safely depend on a small number of frontier providers for cost, access, auditability, and strategic autonomy. The practical direction is orchestration rather than either-or substitution: models, static and dynamic analysis, fuzzing, symbolic execution, and semantic representations should be composed into workflows whose output is measured by validated security impact, not by candidate volume alone.

Throughout the paper, we use \emph{candidate report} to mean a model- or system-generated suspicion that a security-relevant defect may exist. We use \emph{accepted vulnerability} to mean a finding confirmed by the maintainer or responsible security team as a genuine security issue. We use \emph{exploitable vulnerability} for an accepted issue with a credible attack path under an explicit threat model. We use \emph{exploit chain} for a composition of vulnerabilities, techniques, and delivery assumptions that produces access or control under operational constraints. We use \emph{actionable security outcome} more broadly: an accepted report, a minimal reproducer, a proof of concept, a patch, a regression test, or a hardening change that reduces future security risk. This terminology is deliberately conservative because much of the public debate collapses these categories into a single word, ``bug,'' and then treats all reported counts as economically equivalent.

Recent practitioner guidance has framed Mythos as the emblem of an ``AI Vulnerability Storm,'' arguing that Mythos-class systems will increase the speed and volume of vulnerability discovery, exploit generation, patch pressure, and automated attacks~\cite{csa_mythos_ready}. We agree that LLM-assisted vulnerability workflows change bug economics, but the economically important development is wider and more precise than a single restricted frontier model. LLM-assisted workflows existed before Mythos~\cite{bynario-idea,aixcc-results}; frontier models are becoming more capable; open-weight models are becoming more usable; and effective harnesses increasingly encode the security workflow around them. The defender question is not simply whether a model can find a bug. It is whether the model changes the cost of producing validated findings, proofs of impact, patches, and review packages faster than organizations can triage and ship them.

In summary, we make four contributions:

\begin{itemize}
\item First, we reframe \emph{bugonomics} as a shift from a public high-end narrative dominated by zero-day asymmetry toward broader defender-side remediation throughput, rather than as model-inference accounting.
\item Second, we connect public exploit-market prices, vulnerability reward programs, incident baselines, and Mythos/Firefox data to show why LLMs reprice defender workflows without making all generated bugs operationally valuable.
\item Third, we introduce a simple cost model that separates candidate generation, validated finding production, impact evidence, remediation packaging, and maintainer triage.
\item Fourth, we separate the economics of defender-led hardening from the economics of unconstrained zero-day hunting, showing why candidate reports, impact evidence, remediation packages, and operational exploit chains should not be treated as interchangeable outputs.
\end{itemize}

\section{Background and Related Work}\label{sec:background}
\subsection{Automated Vulnerability Discovery}
Automated vulnerability discovery predates LLMs by decades. Static analysis, fuzzing, symbolic execution, and hybrid testing systems have all improved the ability to find bug classes at scale. AFL-style fuzzing popularized coverage-guided mutation as a practical engine for discovering memory-safety bugs \cite{afl}; libFuzzer and sanitizer-based workflows brought similar ideas into continuous testing \cite{libfuzzer,asan}; symbolic execution systems such as KLEE explored program paths with constraint solving \cite{klee}. These systems changed bugonomics before LLMs did: they lowered the cost of finding certain classes of defects, especially when crashes, traces, counterexamples, or sanitizer failures could provide executable evidence.

They changed zero-day asymmetry without eliminating it. A crashing input is not necessarily a useful primitive, a useful primitive is not necessarily a reliable exploit, and a reliable exploit against a test configuration is not necessarily an operational chain against a hardened target. This gap helps explain why high-end vulnerability research remained expensive even after fuzzing and static analysis matured: tooling reduced search cost for some bug classes, but vulnerability value still depended on exploitability, chaining, stealth, reliability, and deployment context.

The distinction between defect existence and operational exploitability is not new. Dullien's weird-machines framing treats exploitation as learning to control emergent computation created by a defect and its execution environment; a bug matters operationally only when it yields useful primitives under realistic constraints \cite{dullien-weird-machines}. That distinction underlies the artifact categories used throughout this paper: candidate report, accepted vulnerability, impact-backed finding, remediation package, and production exploit chain.

LLM-assisted vulnerability workflows change the economics differently. Their value is not only that models can summarize code or suggest bug patterns, but that they can be embedded in harnesses that retrieve context, generate tests, run tools, deduplicate results, produce proofs of impact, and prepare reports or patches across many targets in parallel. The relevant capability is therefore not a standalone model judgment that a bug exists; it is the ability of an LLM-driven pipeline to turn codebase context and tool feedback into evidence that a defender or maintainer can act on.

\subsection{Offensive Pricing Before LLMs}
Before LLMs, the clearest public bugonomics signal came from offensive pricing. The private market prices operational capability, not merely novelty. A zero-day vulnerability may be valuable because no patch exists, but the highest prices attach to exploit chains that deliver a result under constraints: remote reachability, no user interaction, sandbox escape, persistence, supported target versions, reliability, and low detection risk. Public price lists and reporting are imperfect market windows, but they show the order of magnitude. In 2024, Crowdfense's public acquisition program reportedly offered \$5--\$7 million for iPhone zero-days, up to \$5 million for Android, up to \$3 million for Chrome, up to \$3.5 million for Safari, and \$3--\$5 million for WhatsApp or iMessage \cite{techcrunch-zero-day-prices}. These are not prices for ``a bug'' in the abstract. They are prices for exclusive, working capability.

Those prices also show why the public high-end bugonomics narrative was dominated by attackers and offensive buyers. When platforms become harder to exploit, prices rise; when a class is oversupplied, prices can fall. The same TechCrunch report describes prices multiplying as major vendors hardened products, and quotes industry participants attributing the increase to greater complexity and time required for exploitation \cite{techcrunch-zero-day-prices}. RAND's study provides a complementary view: zero-day vulnerabilities can survive for years, but their policy significance comes from their usefulness as offensive, defensive, and research assets \cite{rand-zero-days}. The economic object is therefore not merely the defect's existence. It is the asset created by knowing, validating, exploiting, withholding, disclosing, or fixing the defect.

This historical baseline matters because LLMs change who can economically participate. If models reduce the cost of finding candidate defects, writing reproducers, generating proofs of impact, and preparing patches, defenders can turn vulnerability discovery into a throughput problem rather than a boutique exploit-acquisition problem. The exploit does not disappear from the defender workflow; it changes role. It becomes evidence of impact and a prioritization tool, not necessarily the final commodity being bought or sold.

\subsection{Defensive Repricing, Incident Baselines, and LLMs}
Public vulnerability reward programs provide the defender-side price signal. They are not equivalent to offensive exploit markets: a bounty usually rewards responsible disclosure and evidence sufficient for a vendor to act, not exclusive access or operational stealth. Still, they show what large defenders are willing to buy. Google's VRP reported more than \$17 million in 2025 payouts, an all-time high and more than a 40\% increase over 2024 \cite{google-vrp-2025}. Google's later Android and Chrome VRP update explicitly adjusted rewards and bonuses around high-value categories in the AI era \cite{google-vrp-ai-era}. This is bugonomics in practice: defenders adjust incentives when the expected supply and quality of reports change.

This defender-side repricing did not begin with LLMs. Google Project Zero is an important pre-LLM example: it uses offensive-grade research, public disclosure, exploit tracking, and root-cause analysis to make zero-days harder to use at scale \cite{projectzero-in-the-wild,projectzero-rca}. Mature VRPs are another example: they create legal markets that pay for reports, reproduction, and security value rather than for exclusive operational access. LLM-assisted systems therefore do not invent defensive bugonomics. They change its scale and distribution: activities previously concentrated in well-funded vendors, elite research teams, and mature bounty programs can increasingly be automated, repeated, and applied across large codebases.

Incident data provides a third baseline. Verizon's 2025 DBIR reports vulnerability exploitation as an initial access vector in 20\% of breaches, close to credential abuse at 22\% and above phishing at 16\% \cite{verizon-dbir}. Mandiant's M-Trends 2025 reports exploitation as the most frequently observed initial infection vector in its incident-response investigations, at 33\% where a vector was identified, followed by stolen credentials at 16\% and email phishing at 14\% \cite{mandiant}. Google Threat Intelligence Group tracked 90 zero-days exploited in the wild in 2025, with enterprise technologies making up 48\% of the total \cite{google-zero}. These figures show that exploitation matters materially, but they also show why ``cheaper bug finding'' does not map linearly to attacker capability. Attackers substitute across paths: known vulnerabilities, credentials, exposed edge systems, social engineering, insiders, and supply-chain access can be cheaper than bespoke exploit development.

The recent Mythos debate should be read against those market signals. Anthropic and Mozilla report impressive frontier-model results, but the economically important lesson is not that one model name replaces the exploit market. It is that LLM-assisted systems can reprice several stages of defender work: candidate generation, context retrieval, harness generation, dynamic test-case generation, deduplication, proof-of-impact construction, report drafting, patch suggestion, and release integration. Mozilla's Firefox 150 account is especially useful because it describes an agentic harness built on fuzzing infrastructure, dynamic test-case generation, parallel execution, deduplication, bug tracking, triage, patch validation, and release integration \cite{mozilla-hacks-firefox150}. That is the bridge to the model in the next section: bugonomics should ask which stage is being made cheaper, which defensive bottleneck grows, and whether the output is a candidate report, a validated finding, an impact proof, or a shipped fix.

\section{A Bugonomics Model}\label{sec:cost}
Bugonomics must separate production cost from market price. The cost to produce a candidate, finding, proof of impact, or fix is not the same as the price a buyer will pay for it. Price also reflects buyer utility, scarcity, exclusivity, operational timing, legal risk, substitutability, and the availability of cheaper paths to the same objective. We therefore do not infer private exploit-market prices from public LLM reports. Instead, we use offensive prices as one historical baseline and model which defender-side production stages LLM-assisted systems plausibly make cheaper.

Let $P_o$ be the price or value of an outcome $o$, where $o$ may be a candidate report, a validated finding, a proof of impact, or a remediation package. At a high level,
\begin{equation}
P_o = f(U_o,S_o,X_o,R_o,A_o),
\label{eq:price}
\end{equation}
where $U_o$ is buyer utility, $S_o$ is scarcity, $X_o$ captures exclusivity and operational constraints, $R_o$ captures legal and disclosure risk, and $A_o$ captures available substitutes. Production cost constrains supply, but it does not determine price by itself. This is why a cheap candidate report, a validated high-severity bug, and a proof of impact can all refer to the same underlying defect while occupying different economic categories.

On the production side, we model LLM-driven defensive vulnerability work as a pipeline that begins with candidate generation but ends only when the defender or maintainer receives an artifact they can act on. Let $C_G$ denote the cost of generating candidate reports, $C_V$ the cost of validating, reproducing, and deduplicating those reports, $C_I$ the cost of establishing impact or exploitability when required for prioritization, $C_R$ the cost of packaging remediation evidence such as tests, patches, and review rationale, and $C_T$ the cost of prioritizing, disclosing, reviewing, and shipping remediation. The total campaign cost is:
\begin{equation}
C_{total}=C_G+C_V+C_I+C_R+C_T.
\label{eq:total}
\end{equation}

For token-based model use, a first-order candidate-generation cost can be expressed as:
\begin{equation}
C_G = q \cdot (T_{in} r_{in} + T_{out} r_{out}) + C_{tools},
\label{eq:generation}
\end{equation}
where $q$ is the number of runs, $T_{in}$ and $T_{out}$ are average input and output tokens per run, $r_{in}$ and $r_{out}$ are dollar prices per token, and $C_{tools}$ captures non-token costs such as build infrastructure, sandbox execution, indexing, and storage. Current Claude API prices range from inexpensive Haiku models to Opus-class models; for example, Anthropic lists Opus 4.6 at \$5 per million input tokens and \$25 per million output tokens, Sonnet 4.6 at \$3 and \$15, and Haiku 4.5 at \$1 and \$5 \cite{anthropic-pricing}. Batch processing and prompt caching can reduce costs; fast-mode and data-residency options can increase them.

However, the economically relevant denominator is not candidate findings. Let $N_c$ be candidate reports, $\pi_s$ the fraction confirmed as genuine security vulnerabilities, and $\pi_e$ the fraction of those that are exploitable under the relevant threat model. At the campaign level, the cost of producing validated findings is:
\begin{equation}
C_F = C_G + C_V,
\label{eq:finding_cost}
\end{equation}
and the corresponding cost per validated finding before recipient-side triage is:
\begin{equation}
C_{finding}=\frac{C_G+C_V}{N_c \cdot \pi_s}.
\label{eq:finding_unit}
\end{equation}
If the desired outcome is an impact-backed finding rather than a validated report, the numerator must also include proof-of-impact work:
\begin{equation}
C_{impact}=\frac{C_G+C_V+C_I}{N_c \cdot \pi_s \cdot \pi_e}.
\label{eq:impact_outcome}
\end{equation}
If the desired outcome is a maintainer-accepted report, accepted fix, or remediation package, remediation packaging and maintainer-side triage and review costs must also be included:
\begin{equation}
C_{accepted}=\frac{C_G+C_V+C_R+C_T}{N_c \cdot \pi_s}.
\label{eq:accepted}
\end{equation}
This distinction matters because public discussion often collapses candidate generation, validated vulnerability discovery, impact proof, and remediation into a single notion of ``finding bugs.''

Validation cost can be approximated as:
\begin{equation}
C_V = N_c \cdot h_v \cdot w_v,
\label{eq:validation}
\end{equation}
where $h_v$ is validation time per candidate and $w_v$ is the fully loaded hourly cost of the validator. In this model, validation is not a downstream activity after a bug has already been found; it is part of producing the finding. Impact assessment can be approximated as:
\begin{equation}
C_I = N_s \cdot h_i \cdot w_i,
\label{eq:impact}
\end{equation}
where $h_i$ is impact or exploitability-assessment time per accepted finding, $w_i$ is the corresponding hourly cost, and $N_s=N_c\pi_s$. Remediation packaging can be approximated as:
\begin{equation}
C_R = N_s \cdot h_r \cdot w_r,
\label{eq:remediation_package}
\end{equation}
where $h_r$ is the time to produce tests, patches, review rationale, and other remediation artifacts per accepted finding, and $w_r$ is the corresponding hourly cost. This term is intentionally separate from $C_I$: proving that a bug matters is not the same as preparing a change that maintainers can safely ship.

Triage cost applies to accepted findings that enter disclosure, prioritization, or remediation:
\begin{equation}
C_T = N_s \cdot h_t \cdot w_t,
\label{eq:triage}
\end{equation}
where $h_t$ is maintainer or recipient triage time per accepted finding and $w_t$ is the corresponding hourly cost. These equations are workflow accounting abstractions. They do not model supply, demand, price elasticity, or market clearing. Their purpose is to make hidden work visible: false positives, exploitability uncertainty, remediation packaging, and recipient review burden.

Semantic grounding\footnote{Here, semantics grounding refers to vulnerability reasoning aligned with program behavior, execution constraints, attacker control, and exploitability conditions.} affects these terms without needing to be a separate output category. Evidence tied to program behavior, executable traces, attacker control, path feasibility, privilege boundaries, reachable configurations, or exploitability assumptions can reduce $h_v$, $h_i$, and $h_t$. Conversely, a system that relies on shallow proxies can inflate $N_c$ without proportionally increasing $N_s$, raising validation and triage cost even when candidate generation is cheap.

\section{What the Public Numbers Actually Support}\label{sec:numbers}
\subsection{Mythos Preview}
A common shorthand is that Mythos found a ``\$20,000 bug.'' That is not what the public Anthropic post strictly says. The post states that, across a thousand runs through Anthropic's scaffold, the total cost was under \$20,000 and the campaign found several dozen more findings. It also states that the specific run that found the showcased OpenBSD bug cost under \$50, but that this number only makes sense with hindsight because one cannot know in advance which run will succeed \cite{anthropic-mythos}.

The defensible arithmetic is therefore campaign-level, not per-bug. If ``several dozen'' is interpreted conservatively as 24--48 findings, the token/scaffold cost per reported finding is bounded above by approximately \$417--\$833:
\begin{equation}
\frac{\$20{,}000}{48} \leq C_{\mathrm{reported}} \leq \frac{\$20{,}000}{24}.
\end{equation}
If the number of findings is higher, the average is lower. If some findings are duplicates, low severity, or not security-relevant, the cost per actionable vulnerability is higher. The public text does not provide enough information to determine $\pi_s$, $\pi_e$, validation hours, deduplication effort, or maintainer cost. The rigorous conclusion is therefore not ``\$20,000 per bug.'' It is that candidate generation can be surprisingly cheap in API terms, especially for a frontier lab operating on its own infrastructure, while total actionable cost remains under-specified.

\subsection{Firefox Collaboration}
The Firefox collaboration gives a more useful precision anchor. Anthropic reports that Claude Opus 4.6 discovered 22 Firefox vulnerabilities over two weeks, 14 assigned high severity by Mozilla, after scanning nearly 6,000 C++ files and submitting 112 unique reports \cite{anthropic-mozilla}. This yields:
\begin{equation}
\pi_s = \frac{22}{112}=19.6\%, \quad \pi_h=\frac{14}{112}=12.5\%,
\end{equation}
where $\pi_h$ is the high-severity acceptance fraction among submitted reports. Equivalently, the campaign required about 5.1 submitted reports per accepted vulnerability and 8 submitted reports per high-severity vulnerability.
This reviewer burden is economically important because downstream validation and maintainer attention may become the dominant constraint as candidate generation scales.

The public post does not disclose the API cost of the Firefox candidate-generation campaign. The useful public signal is instead the scale of filtering and the type of lifecycle work required. For orientation only, if a campaign of similar scale cost between \$5,000 and \$20,000 in model and scaffold expenditure, candidate-generation cost alone would be \$227--\$909 per accepted vulnerability and \$357--\$1,429 per high-severity vulnerability. If validation requires 0.5--2 hours per submitted report at a fully loaded expert cost of \$100--\$250/hour, then validation adds:
\begin{equation}
112 \cdot (0.5\text{--}2) \cdot (100\text{--}250)=\$5{,}600\text{--}\$56{,}000.
\end{equation}
Amortized over 22 accepted vulnerabilities, that is \$255--\$2,545 per accepted vulnerability; amortized over 14 high-severity vulnerabilities, it is \$400--\$4,000 per high-severity vulnerability. This arithmetic is not a direct measurement of the Firefox campaign. It is a sensitivity check showing that validation can become the same order of magnitude as disclosed model or scaffold spending once report volume exists. Impact assessment, exploit development, remediation packaging, and maintainer-side patch review would add further cost that the public reports do not quantify.

Mozilla's later account of the Firefox 150 hardening effort provides a stronger pipeline anchor. Mozilla reported 271 Mythos-identified bugs for Firefox 150, including 180 sec-high, 80 sec-moderate, and 11 sec-low issues; it also reported 423 total Firefox security bugs fixed across sources in April 2026 \cite{mozilla-hacks-firefox150}. The important detail for bugonomics is that this was not a model-only scan. Mozilla describes an agentic harness built on existing fuzzing infrastructure, dynamic test-case generation, parallel execution on ephemeral VMs, deduplication against known issues, bug tracking, triage, patch validation, and release integration. The discovery subsystem was necessary, but the full security bug lifecycle was what made the output useful at release scale.

\begin{table}[t]
\caption{Public Anchors and Illustrative Derived Quantities}
\centering
\begin{tabular}{p{0.47\linewidth}r}
\toprule
\textbf{Quantity} & \textbf{Value} \\
\midrule
Mythos campaign cost & $<\$20{,}000$ \\
Mythos campaign size & 1,000 scaffolded runs \\
Mythos findings & ``several dozen'' \\
Implied cost/finding if 24--48 findings & $<\$417$--$\$833$ \\
Firefox submitted reports & 112 \\
Firefox accepted vulnerabilities & 22 \\
Firefox high-severity vulnerabilities & 14 \\
Firefox accepted fraction & 19.6\% \\
Firefox high-severity fraction & 12.5\% \\
Reports per accepted vulnerability & 5.1 \\
Reports per high-severity vulnerability & 8.0 \\
Firefox 150 Mythos-identified bugs & 271 \\
Firefox 150 severity split & 180 high, 80 moderate, 11 low \\
Firefox April 2026 total security fixes & 423 \\
Exploit-development experiment & $\sim\$4{,}000$ for 2 crude exploits \\
\bottomrule
\end{tabular}
\label{tab:public_numbers}
\end{table}

The same Anthropic report states that several hundred exploit-development attempts cost approximately \$4,000 and produced two crude browser exploits in a testing environment with security features removed \cite{anthropic-mozilla}. This implies roughly \$2,000 in API credits per successful crude exploit in that artificial setting. It does not imply \$2,000 per production exploit, and it does not imply sandbox escape. Anthropic explicitly notes that Firefox's defense-in-depth would have mitigated those particular exploits. This distinction is central to bugonomics: discovery, exploitability, and operational impact are different economic objects.

\subsection{Why True-Positive Replay Is Not Production Discovery}
Replicating known true positives is useful as a regression test, but it is not a production-discovery metric. Once the target bug, affected component, vulnerable pattern, or expected exploit primitive is known, the search problem has already been simplified. The economically relevant question is not whether a system can reproduce a known finding, but how much time, inference, tool execution, and harness engineering are required before it reaches that region of the search space without being guided there.

This matters because vulnerability discovery is path-dependent. A model may have the latent capability to recognize a bug once enough context is assembled, yet still require many failed runs before the right files, call paths, input assumptions, and threat model align. In that setting, the important quantity is not binary capability but time-to-first-useful-finding at a given budget. A system that finds a known vulnerability after being pointed at the right subsystem is not economically equivalent to one that discovers the same vulnerability during an unguided campaign over a large codebase.

Scaffolding complicates the interpretation further. Engineering around a known bug class can make a system look stronger while narrowing what it is actually able to discover. This is valuable when the goal is systematic hardening against a known class: for example, generating harnesses, sanitizers, queries, or validators for a recurring memory-safety pattern. But it should not be confused with general vulnerability discovery. In that case, part of the capability has moved from the model into the scaffold, and the real measurement question becomes which bug classes the scaffold supports, how much engineering was required to support them, and how often the resulting pipeline generalizes.

The problem is sharper for logic bugs. Authorization errors, parser-state mistakes, protocol-state violations, business-invariant failures, and cross-component trust-boundary bugs often do not reduce to a reusable syntactic signature. They depend on application-specific semantics: what state should be impossible, which actor is allowed to trigger a transition, what compatibility promise must be preserved, and which data was supposed to remain untrusted. Pure engineering can still help by extracting traces, invariants, call graphs, and tests, but supporting a logic-bug class is less straightforward than supporting a sanitizer-detectable memory-safety class. High variance is therefore not an incidental detail; it is part of the cost model.

\section{What Future Reports Should Measure}\label{sec:measurement}
The public numbers are useful, but they are not sufficient for comparing systems or estimating operational impact. A vulnerability-discovery report that discloses only candidate counts and a few examples can demonstrate capability, but it cannot establish cost effectiveness. A report that discloses only confirmed vulnerabilities can demonstrate impact, but it hides the amount of failed search, deduplication, and maintainer attention required to get there. Bugonomics requires both numerators and denominators.

We therefore recommend that future LLM-assisted vulnerability-discovery reports separate raw generation from validated outcomes. At minimum, reports should distinguish raw candidates, deduplicated candidates, submitted reports, accepted findings, severity distribution, exploitability status, grounding evidence, validation time, maintainer review time, patch status, time-to-first-useful-finding, run count, failed-run cost, scaffold engineering effort, and total compute or API expenditure. These quantities need not disclose sensitive exploit details or private prompts. They can be aggregated at campaign level while still making the economic interpretation substantially clearer.

\begin{table}[t]
\caption{Minimum Reporting Fields for LLM-Assisted Vulnerability Campaigns}
\centering
\begin{tabular}{p{0.43\linewidth}p{0.45\linewidth}}
\toprule
\textbf{Field} & \textbf{Why it matters} \\
\midrule
Raw candidate count & Measures model or harness output volume before filtering. \\
Deduplicated candidate count & Separates repeated discoveries from distinct hypotheses. \\
Submitted report count & Captures what the maintainer or program actually had to review. \\
Accepted finding count & Measures confirmed security value rather than generated text. \\
Precision / reviewer burden & Captures how many candidate findings must be reviewed per accepted security outcome.\\
Severity distribution & Distinguishes low-risk cleanup from high-impact vulnerabilities. \\
Impact/exploitability status & Separates suspected defects from credible attack paths. \\
Grounding evidence & Records reproducers, traces, path constraints, attacker-control arguments, and affected configurations. \\
Validation hours & Exposes human or automated confirmation cost. \\
Impact-assessment hours & Captures the cost of proving severity or exploitability. \\
Maintainer review hours & Captures the scarce resource most likely to bottleneck open source. \\
Patch and test status & Indicates whether the finding became a durable security improvement. \\
Time-to-first-useful-finding & Captures search cost rather than replay success. \\
Run count and failed-run cost & Measures variance and unsuccessful exploration. \\
Scaffold engineering effort & Separates model capability from bug-class-specific support. \\
Compute/API/tooling cost & Enables comparison across frontier, open-weight, and hybrid systems. \\
\bottomrule
\end{tabular}
\label{tab:reporting_fields}
\end{table}

The most important derived quantities then become straightforward: precision among submitted reports, cost per accepted vulnerability, cost per high-severity vulnerability, cost per exploitable vulnerability, and accepted fixes per maintainer-hour consumed. These metrics also make it harder to confuse different claims. A system may be excellent at generating hypotheses, mediocre at exploitability assessment, and still useful for hardening campaigns if it produces cheap reproducible tests. Another system may produce fewer reports but a higher accepted-finding rate, making it more suitable for maintainers with limited review capacity. Without the fields in Table~\ref{tab:reporting_fields}, public claims about ``AI-discovered vulnerabilities'' remain difficult to compare.

\section{Reward Programs as Market Repricing}\label{sec:rewards}
Vulnerability reward programs (VRPs) are one of the few public markets where changes in bug economics become visible. They do not measure the full value of a vulnerability, and bounty prices are not the same thing as the cost of producing validated findings. But they do reveal what large software vendors are willing to reward when external researchers submit findings. As AI-assisted systems lower the marginal cost of candidate generation, reward programs have an incentive to distinguish more sharply between low-signal reports and reports that reduce real remediation cost.

Google's recent Android and Chrome VRPs update is a useful signal. The announcement describes immediate changes to reward amounts and bonus structures intended to reflect the report types and bug categories that provide the most security value today \cite{google-vrp-ai-era}. The important point is not that Google has declared a new universal price for AI-discovered bugs. It is that a major buyer of external vulnerability research is explicitly adjusting incentives in response to a changing research environment. That is bugonomics in practice: when candidate production becomes cheaper, the market reprices validation, depth, exploitability, and usefulness.

This suggests a likely direction for mature VRPs. Raw candidate volume should become less valuable. Reports that include reliable reproduction, affected-version analysis, exploitability reasoning, minimized test cases, patch guidance, regression tests, or documentation of changed behavior should become more valuable because they reduce the recipient's validation and triage burden. In the terminology of this paper, reward programs will increasingly pay not for generated suspicion but for review-package quality.

The offensive-market effect is secondary to this paper's defender-centered argument. Higher defensive rewards can improve the opportunity cost of responsible disclosure, but they do not erase the value of stealth, exclusivity, exploit-chain composition, target specificity, and operational timing. As candidate-generation cost falls, legal markets may absorb more commodity findings and better-packaged high-signal reports, while private offensive markets continue to value vulnerabilities that produce reliable access under real constraints. The likely future is therefore not a single price collapse. It is stratification: low-signal candidate reports become cheaper, validated remediation packages become a premium defensive good, and production-grade exploit chains remain economically distinct.

\section{Incentives, Telemetry, and Open Source Pressure}\label{sec:incentives}
\subsection{Offensive Choices Are Economic Choices}
The apocalyptic framing assumes that reducing the cost of novel vulnerability discovery directly and linearly increases attacker capability. The incident baseline is more complex. Verizon reports vulnerability exploitation in 20\% of breaches, close to credential abuse at 22\% and above phishing at 16\% \cite{verizon-dbir}. Mandiant, in a different incident-response population, reports exploitation as the most common identified initial infection vector at 33\%, followed by stolen credentials and phishing \cite{mandiant}. Google tracked 90 zero-days exploited in the wild in 2025 and expects AI to accelerate reconnaissance, vulnerability discovery, and exploit development \cite{google-zero}.

Three implications follow. First, exploitation is economically important and cannot be dismissed. Second, most exploitation observed in enterprise incidents is not necessarily novel zero-day exploitation; known vulnerabilities, patch latency, edge devices, VPNs, and exposed enterprise software remain central. Third, the defender value of AI may be highest where it compresses time-to-remediation: finding bug classes before adversaries exploit them, generating reproducible test cases, producing candidate patches, and prioritizing known exposed weaknesses.

\subsection{Bug Age Is Not a Quality Metric}
Recent vulnerability-discovery narratives often emphasize that a model found a bug that had been present for ten, sixteen, or twenty years. This is an interesting observable fact, but it is a weak security metric. Age does not measure exploitability, reachability, primitive quality, exploit-chain utility, severity, or discovery difficulty. A twenty-year-old bounds-check bug in unreachable code and a two-week-old remotely reachable memory-corruption primitive are not comparable simply because one is older.

The more important point is historical. For more than two decades, offensive research groups, boutique exploit labs, government contractors, brokers, and independent researchers have searched for vulnerabilities in precisely these mature codebases. The fact that a bug is old does not imply that it was unknown. It only implies that it was unpatched in the public codebase. In some cases, unusually rapid publication of technical details after a patch has historically functioned as a signal that others likely had the bug or enough adjacent knowledge to weaponize it. Public disclosure records are therefore not a complete record of private vulnerability knowledge.

This matters for bugonomics because age is often used as a proxy for model capability. It should not be. A model finding an old bug may be impressive, but the age alone contributes almost no information about the quality of the finding. The relevant variables are whether the bug is attacker-reachable, whether it provides a useful primitive, whether it composes with other bugs, whether mitigations block exploitation, and how much validation and exploitability analysis are required to turn the candidate into an actionable security finding.

In the notation of Section~\ref{sec:cost}, bug age does not directly affect $C_G$, $C_V$, $C_I$, or $C_T$. It may correlate with code maturity or neglected attack surface, but it does not by itself lower validation cost or increase exploitability probability. Treating age as a capability metric therefore turns an observable property of a defect into a marketing-friendly proxy for discovery difficulty, exploitability, and security value.

\subsection{Why Zero-Days Are Rarely Visible in Commodity Intrusions}
A second recurring mistake is to infer from public incident data that zero-days and high-end vulnerability research are not useful in real attacks. The better explanation is economic. Offensive vulnerability research is expensive, uncertain, and operationally demanding. A team may spend months without finding a suitable bug; may find a bug that lacks the right primitive; may obtain a primitive that does not compose into a reliable exploit; or may build an exploit that is neutralized by target-specific mitigations. Each failure mode consumes time and capital.

Attackers therefore substitute. If the goal is to compromise a package registry, a build system, or a developer account, buying stolen credentials, recruiting insiders, phishing maintainers, compromising CI tokens, or constructing a fake company may be cheaper and more reliable than buying or developing a remote zero-day against the infrastructure itself. This is not evidence that zero-days lack value. It is evidence that attackers optimize across available paths.

The public record is also biased. High-end vulnerabilities are often used by actors and against targets that do not generate commodity breach telemetry. Some exploits are burned quietly; some are patched without full public reconstruction; some are inferred only from timing, patch detail, or private community knowledge. Incident reports therefore undercount parts of the offensive market by design. Public telemetry is useful for estimating common intrusion paths, but it should not be mistaken for a complete map of private vulnerability use.

The implication for this paper is that both extremes are wrong. The fact that credential abuse and known vulnerabilities dominate many enterprise datasets does not make zero-days irrelevant. Conversely, the existence of valuable zero-day markets does not imply that cheaper candidate generation automatically transforms all attacker economics. The missing variable is substitution: attackers choose the cheapest reliable path to their objective.

The forward-looking question is therefore not only whether LLMs increase the supply of zero-days, but how they reshape the timing, scale, and economics of exploitation workflows more broadly. VulnCheck's 2026 State of Exploitation report identified 884 known exploited vulnerabilities with first-time exploitation evidence in 2025, and found that 28.96\% were exploited on or before the day their CVE was published \cite{vulncheck-state-2026}. Its 2026 Exploit Intelligence Report also tracked more than 14,000 exploits for over 10,000 CVE-2025 vulnerabilities, while noting that AI-generated proof-of-concept code is increasingly polluting risk assessment pipelines and that only about 1\% of 2025 CVEs were confirmed exploited in the wild by year end \cite{vulncheck-veir-2026}.

For bugonomics, the important point is not simply whether every vulnerability becomes exploitable, but where scarcity remains in the pipeline as automation improves. Today, prioritization, exploitability assessment, validation, and remediation capacity remain important constraints. Future systems may compress some of these stages substantially. The broader economic shift is therefore toward faster and more scalable vulnerability reasoning, exploit generation, and operationalization, potentially reducing defender reaction time even when the underlying vulnerabilities are already public.

\section{Open Source and the Cost of Public Remediation}\label{sec:opensource}
Open source is where the bugonomics problem becomes most visible. LLM-assisted discovery can increase the supply of candidate reports against widely used projects, but it does not automatically increase the supply of maintainer attention, validation capacity, release engineering, or long-term funding. As $C_G$ falls, the limiting cost may concentrate in $C_V$, $C_I$, $C_R$, and $C_T$, with maintainers absorbing much of the burden in unpaid hours rather than in invoices.

That time burden is not clerical. A plausible vulnerability report can force a maintainer to reconstruct unfamiliar code paths, determine whether an input is attacker-controlled, check affected versions and build configurations, reproduce or reject a proof of concept, evaluate patch side effects, coordinate a release, and communicate with downstream consumers. Even a false positive can be expensive if it consumes scarce expert attention. For a commercial security team, those hours can be budgeted. For many open-source projects, they are taken from nights, weekends, and already-overcommitted maintenance time.

This is why the common framing of open source as being overwhelmed by ``AI slop'' is incomplete. Low-quality reports are a real problem, but the deeper issue is incentive alignment. Many companies depend critically on open-source software while contributing unevenly to the cost of understanding, hardening, and maintaining it. If interest in AI-assisted open-source security is driven mainly by marketing cycles, the result will be temporary attention followed by the same chronic underfunding, but with a new asymmetry: attackers and well-funded defenders retain cheaper discovery capability, while many open-source maintainers lack the budget to run comparable pipelines themselves and must absorb a larger stream of externally generated reports to validate, prioritize, fix, and ship.

A more durable model is defender-funded autonomous software understanding. Organizations that depend on open-source components have direct incentives to run LLM-assisted, tool-verified analysis against the software they rely on, produce high-quality reports and patches, and upstream the results.
For that model to work, the discovering organization must bear not only $C_G$ but also $C_V$ and $C_R$ before disclosure, delivering validated findings with reproducers and candidate patches rather than raw candidates. That shifts the externality back to the funded actor and gives maintainers a $C_T$ they can absorb within existing capacity.
In that model, companies receive proactive security and reduced dependency risk; maintainers receive validated fixes rather than raw candidate reports. The economic target is not report volume, but reduced maintainer cost per accepted fix.

For open-source projects, the relevant metric should therefore be not candidates generated per dollar, but accepted fixes per maintainer-hour consumed.

\section{The Underappreciated Case: Technical Debt}\label{sec:debt}
A high-leverage defender case is technical debt remediation and systematic hardening, not because it is universally more valuable than zero-day hunting, but because the output maps directly to work defenders can ship. Large codebases accumulate deprecated APIs, inconsistent input validation, unchecked error paths, unsafe language boundaries, memory-management hazards, and fragile assumptions around serialization, parsing, and privilege separation. Many of these issues are not individually headline-worthy, but they increase the attack surface and slow secure maintenance.

Technical debt remediation has different economics from unconstrained zero-day discovery. A single analysis pass can surface many actionable issues, and the cost can be amortized across thousands of files and many patches. Maintainers also have advantages external attackers lack: full build systems, test suites, deployment context, code ownership, and authority to ship fixes. This shifts value capture toward defenders and maintainers without requiring every finding to become an operational exploit.

This reframing also clarifies why validation artifacts are valuable. Anthropic reports that Mozilla valued minimal test cases, detailed proofs of concept, and candidate patches \cite{anthropic-mozilla}. These artifacts convert a model-generated suspicion into maintainable engineering work. In bugonomics terms, they reduce $h_v$ and $h_t$ by lowering the cognitive burden on maintainers.

Mozilla's later account adds a second measurement angle: failed search trajectories. While auditing harness logs, Firefox engineers observed repeated model attempts to pursue prototype-pollution escape paths that were blocked by an earlier architectural change to freeze privileged prototypes by default \cite{mozilla-hacks-firefox150}. For bugonomics, this makes failed model behavior useful telemetry: it can show which exploit strategies remain attractive to the search process, and which prior hardening investments are constraining them.

\section{Model Access and Defensive Diffusion}\label{sec:access}
\subsection{Open-Weight Models and Strategic Dependence}\label{sec:openweights}
Open-weight models are central to the economics, but not because every open model must be treated as equivalent to a restricted frontier model. They may reduce direct API cost, avoid metering, support local deployment, and allow inspection or fine-tuning. They also reduce strategic dependence on a small number of frontier providers. Their most credible near-term role is often component-level: first-pass filtering, patch drafting, summarization, harness generation, regression-test generation, or local analysis over private code that cannot be sent to an external API.

At the same time, open weights do not eliminate cost. They replace API cost with compute, engineering, maintenance, evaluation, and validation cost. If a cheaper model produces more candidates with lower precision, it may increase $C_F$ even while reducing $C_G$. Conversely, if a smaller model is used as a first-pass filter or specialized component inside a semantics-grounded harness, it may lower total cost. The correct comparison is therefore not frontier versus open-weight in isolation, but total pipeline cost at a fixed quality target.

\subsection{Access Control and Trusted Defensive Access}
The access-control question is not simply whether a cyber-capable model is public or private. A fully closed model can reduce misuse risk, but it can also make the defensive effect narrow, non-reproducible, and dependent on a vendor's private allocation decisions. A broadly available model can accelerate defenders, but may also lower the cost of offensive experimentation. The interesting design space is between those extremes: verified, trust-based access for defensive users, paired with monitoring, policy constraints, and ecosystem support.

OpenAI's Trusted Access for Cyber is an example of this middle position. OpenAI describes it as an identity- and trust-based framework for placing enhanced cyber capabilities in the hands of defenders, with identity verification for potentially high-risk cybersecurity work, enterprise trusted access paths, and an invite-only track for researchers and teams that need more permissive defensive capability \cite{openai-tac}. OpenAI later described scaling the program to thousands of verified individual defenders and hundreds of teams responsible for critical software, alongside a cyber-permissive GPT-5.4-Cyber variant \cite{openai-tac-scale}. The important contrast is not that one vendor favors open source and another favors large companies. The contrast is between a model whose defensive use is constrained to a small closed circle and a trust-based access model that attempts to broaden defensive availability while preserving safeguards.

For bugonomics, the side effect of complete closure is measurement opacity. If only a small set of actors can run the model, outsiders cannot easily evaluate precision, false-positive rates, validation burden, patch quality, or the kinds of projects that benefit. The output may still be valuable, but the public cannot distinguish whether the model reduced total security work, shifted work to maintainers, or simply generated a private backlog of findings. A trust-scaled program does not solve that problem automatically, but it creates more opportunities for reproducible reporting across different organizations, project types, and validation workflows.

\section{Review Packages Across Code Ownership Models}\label{sec:reviewpackages}
The unit of output for LLM-assisted remediation should not be the patch alone. In large codebases, a security fix may alter business logic, parser behavior, serialization semantics, error handling, compatibility assumptions, privilege boundaries, or internal invariants. One such fix can be reviewed manually. One hundred such fixes create a comprehension problem. A practical vulnerability workflow should therefore produce a \emph{review package}: the patch, the candidate report, the affected code paths, the exploitability argument, the changed invariant, the regression test, the compatibility risk, and any documentation needed to explain the updated behavior.

The content of the review package depends on who, or what, owns the surrounding code. In a fully human-owned codebase, the package should optimize for human review. Maintainers need a concise explanation of the old invariant, why it was unsafe, what the patch changes, how the tests demonstrate the change, and what downstream behavior may break. In this setting, documentation is not decorative. It is part of the fix because it helps future maintainers understand why the code now behaves differently.

In a human-driven LLM codebase, where humans direct development but models draft substantial portions of code, the package must also capture model-facing context. The relevant artifacts include the human intent, the prompt or task specification at a suitable level of abstraction, the model-generated change, the review comments, and the final human decision. The goal is not to preserve every token of interaction, but to preserve enough rationale that a later reviewer can tell whether the fix reflects a deliberate security decision or a model-generated local repair that accidentally changed broader behavior.

In a fully LLM-owned or highly autonomous codebase, the review package becomes closer to an audit trail. The system must explain what objective it optimized, what evidence it used, what alternatives it rejected, what tests it generated, and why the resulting behavioral change is acceptable. Without that package, autonomy can produce patches faster than humans can understand them. The bottleneck then moves from writing code to reconstructing intent after the fact, which is exactly the kind of hidden cost bugonomics is meant to expose. Further, how to trust such a trail is an additional separate open problem.

\begin{table}[t]
\caption{Review Package Emphasis by Code Ownership Model}
\centering
\begin{tabular}{p{0.31\linewidth}p{0.58\linewidth}}
\toprule
\textbf{Ownership model} & \textbf{Review package emphasis} \\
\midrule
Fully human owned & Human-readable rationale, changed invariant, tests, compatibility notes, and documentation updates. \\
Human-driven LLM & Human intent, model-generated diff, review decision, security rationale, and guardrails against local repairs that alter global behavior. \\
Fully LLM-owned & Audit trail, objective specification, evidence used, alternatives rejected, generated tests, and explanation of behavioral acceptability. \\
\bottomrule
\end{tabular}
\label{tab:review_packages}
\end{table}

\section{Discussion: Demystification Without Apocalypse}\label{sec:discussion}
The current evidence supports a measured conclusion. LLM-driven vulnerability discovery is real, improving, and strategically important. It does not follow that defenders are economically obsolete. Public results show that candidate generation is becoming cheaper, but they also expose the centrality of validated finding production, impact assessment, triage, and patching.

The most important empirical indicators to watch over the next six months are not benchmark scores alone. At the campaign level, the reporting fields in Table~\ref{tab:reporting_fields} make precision, cost per accepted vulnerability, cost per high-severity vulnerability, and accepted fixes per maintainer-hour measurable. At the patch level, the review-package fields in Table~\ref{tab:review_packages} make it possible to evaluate whether model-assisted remediation is producing understandable software rather than merely producing diffs. At the ecosystem level, the indicators are broader: (i) the ratio of LLM-attributed CVEs to total CVEs; (ii) changes in precision among submitted reports; (iii) time-to-validation and time-to-patch; (iv) cost per impact-backed finding and accepted remediation package, not only candidate report; (v) the degree to which semantic grounding reduces false positives under distribution shift; and (vi) the degree to which trusted-access and open-weight systems narrow or widen the practical defensive capability gap.

The practical direction is orchestration. Frontier models should be used where their reasoning capability justifies cost. Open-weight models should be used where they provide sufficient capability with better control or economics. Program analysis, fuzzing, symbolic reasoning, dynamic execution, and sanitizer traces should ground and verify model hypotheses.

\section{Threat to Validity}\label{sec:limitations}

This paper relies on public reports whose underlying run logs, prompts, deduplication rules, and maintainer triage records are not available. The Mythos Preview post provides campaign-level cost but not enough detail to compute cost per validated vulnerability. The Firefox collaboration provides candidate and accepted-vulnerability counts, but not API expenditure or maintainer validation time. Any numeric ranges derived from these reports are therefore orientation exercises, not measured facts or the basis of the paper's thesis.
There is also an important visibility bias in the current public record. Most published campaigns come from well-resourced organizations with strong incentives to report successful outcomes, while failed campaigns, excessive false positive rates, abandoned workflows, or unsustainable maintainer burden are far less likely to be disclosed. The publicly visible economics should therefore not be assumed to represent the full distribution of operational outcomes.
Exploit-market price anchors are also imperfect: public broker or acquisition-program prices reveal advertised willingness to pay for certain capabilities, not the full private market. Incident statistics are population-dependent: Verizon and Mandiant sample different kinds of events, and zero-day tracking is affected by detection and disclosure bias. More broadly, model capability and pricing are moving targets; any static cost estimate should be treated as a snapshot. Finally, this paper does not attempt to model the full economics of vulnerability markets. It does not estimate private-market demand, price elasticity, or market-clearing behavior. Its narrower goal is to explain how LLM-assisted systems change the cost structure of producing defender-actionable security artifacts.

\section{Conclusion}\label{sec:conclusion}
LLMs are demystifying parts of vulnerability discovery by lowering the cost of producing plausible, increasingly evidence-backed security findings and by raising the floor of routine security review. They are not, on the public evidence alone, eliminating the economics of exploitability assessment, prioritization, remediation, and maintainer-side operationalization. The reported Mythos and Firefox results are impressive, but the most rigorous reading is that the bottleneck is shifting from mechanical search toward exploitability reasoning, prioritization, remediation, and the production of maintainable security outcomes.

Nor should bug age be mistaken for bug quality. The fact that a defect survived for decades is historically interesting, but it does not establish exploitability, complexity, importance, or novelty to private researchers. Likewise, the relative scarcity of zero-days in public incident datasets should be read through attacker incentives rather than through usefulness. High-end vulnerability research has always been expensive, uncertain, and unevenly visible. Attackers choose cheaper paths when cheaper paths work.

Bugonomics therefore changes the debate. The question is not whether frontier models, open-weight models, or program analysis ``win.'' The question is how to orchestrate them so that scarce validation, prioritization, and release capacity goes toward durable fixes rather than mechanical search and report drafting. A central defender opportunity is technical debt remediation: semantics-grounded, tool-verified, model-assisted workflows that help maintainers find, validate, prioritize, and fix security-relevant defects before they become tomorrow's exploited vulnerabilities.

The defensive opportunity is therefore not simply to celebrate cheaper bug discovery. It is to convert increasingly scalable vulnerability reasoning and evidence generation into accepted fixes, especially in the open-source software that underlies modern infrastructure. The central metric for the next phase of bugonomics should be accepted security improvement per dollar and per maintainer-hour, not candidate reports per model run.

\section*{Acknowledgment}
The authors thank the security research and open-source communities whose public reports, critiques, and disclosures make this discussion possible. In particular, we would like to thank Thomas Dullien, Amjad Fatmi, Kathrin Grosse, Mathias Payer, Kavya Perlam, Konrad Rieck, Krti Tallam, and other anonymous reviewers for their feedback.


\begin{thebibliography}{00}
\bibitem{csa_mythos_ready}
G. Evron, R. T. Lee, R. Mogull, et al.,
``The AI Vulnerability Storm: Building a Mythos-ready Security Program,''
Cloud Security Alliance CISO Community, SANS Institute, [un]prompted,
OWASP Gen AI Security Project, Apr. 18, 2026.
\bibitem{aixcc-results} DARPA, ``AI Cyber Challenge marks pivotal inflection point for cyber defense,'' Aug. 8, 2025. [Online]. Available: \url{https://www.darpa.mil/news/2025/aixcc-results}
\bibitem{bynario-idea} Bynar.io, ``The idea behind BynarIO,'' 2025. [Online]. Available: \url{https://bynar.io/blog/the-idea-behind-bynario}
\bibitem{openai-tac} OpenAI, ``Introducing Trusted Access for Cyber,'' Feb. 5, 2026. [Online]. Available: \url{https://openai.com/index/trusted-access-for-cyber/}
\bibitem{openai-tac-scale} OpenAI, ``Trusted access for the next era of cyber defense,'' Apr. 14, 2026. [Online]. Available: \url{https://openai.com/index/scaling-trusted-access-for-cyber-defense/}
\bibitem{anthropic-mythos} Anthropic Frontier Red Team, ``Assessing Claude Mythos Preview’s cybersecurity capabilities,'' Apr. 2026. [Online]. Available: \url{https://red.anthropic.com/2026/mythos-preview/}
\bibitem{anthropic-mozilla} Anthropic, ``Partnering with Mozilla to improve Firefox's security,'' Mar. 2026. [Online]. Available: \url{https://www.anthropic.com/news/mozilla-firefox-security}
\bibitem{mozilla-hacks-firefox150} B. Grinstead, C. Holler, and F. Braun, ``Behind the Scenes Hardening Firefox with Claude Mythos Preview,'' Mozilla Hacks, May 7, 2026. [Online]. Available: \url{https://hacks.mozilla.org/2026/05/behind-the-scenes-hardening-firefox/}
\bibitem{anthropic-pricing} Anthropic, ``Claude API Pricing,'' 2026. [Online]. Available: \url{https://platform.claude.com/docs/en/about-claude/pricing}
\bibitem{rand-zero-days} L. Ablon and A. Bogart, ``Zero Days, Thousands of Nights: The Life and Times of Zero-Day Vulnerabilities and Their Exploits,'' RAND Corporation, 2017. [Online]. Available: \url{https://www.rand.org/pubs/research_reports/RR1751.html}
\bibitem{techcrunch-zero-day-prices} L. Franceschi-Bicchierai, ``Price of zero-day exploits rises as companies harden products against hackers,'' TechCrunch, Apr. 6, 2024. [Online]. Available: \url{https://techcrunch.com/2024/04/06/price-of-zero-day-exploits-rises-as-companies-harden-products-against-hackers/}
\bibitem{projectzero-in-the-wild} Google Project Zero, ``About 0-days In-the-Wild.'' [Online]. Available: \url{https://googleprojectzero.github.io/0days-in-the-wild/about.html}
\bibitem{projectzero-rca} Google Project Zero, ``Root Cause Analyses,'' 0-days In-the-Wild. [Online]. Available: \url{https://googleprojectzero.github.io/0days-in-the-wild/rca.html}
\bibitem{google-vrp-2025} Google Vulnerability Rewards Program Team, ``VRP 2025 Year in Review,'' Google Security Blog, Mar. 31, 2026. [Online]. Available: \url{https://blog.google/security/vrp-2025-year-in-review/}
\bibitem{google-vrp-ai-era} Google Bug Hunters, ``Evolving the Android \& Chrome VRPs for the AI Era,'' Apr. 30, 2026. [Online]. Available: \url{https://bughunters.google.com/blog/evolving-the-android-chrome-vrps-for-the-ai-era}
\bibitem{verizon-dbir} Verizon, ``2025 Data Breach Investigations Report: Executive Summary,'' 2025. [Online]. Available: \url{https://www.verizon.com/business/resources/reports/2025-dbir-executive-summary.pdf}
\bibitem{mandiant} Mandiant, ``M-Trends 2025,'' 2025. [Online]. Available: \url{https://services.google.com/fh/files/misc/m-trends-2025-en.pdf}
\bibitem{google-zero} Google Threat Intelligence Group, ``Look What You Made Us Patch: 2025 Zero-Days in Review,'' Mar. 2026. [Online]. Available: \url{https://cloud.google.com/blog/topics/threat-intelligence/2025-zero-day-review}
\bibitem{vulncheck-state-2026} P. Garrity, ``VulnCheck State of Exploitation 2026,'' VulnCheck, Jan. 21, 2026. [Online]. Available: \url{https://www.vulncheck.com/blog/state-of-exploitation-2026}
\bibitem{vulncheck-veir-2026} C. Condon, ``Introducing the 2026 VulnCheck Exploit Intelligence Report,'' VulnCheck, Feb. 25, 2026. [Online]. Available: \url{https://www.vulncheck.com/blog/2026-vulncheck-exploit-intelligence-report}
\bibitem{afl} M. Zalewski, ``American Fuzzy Lop,'' 2013. [Online]. Available: \url{https://lcamtuf.coredump.cx/afl/}
\bibitem{libfuzzer} LLVM Project, ``libFuzzer: a library for coverage-guided fuzz testing.'' [Online]. Available: \url{https://llvm.org/docs/LibFuzzer.html}
\bibitem{asan} K. Serebryany, D. Bruening, A. Potapenko, and D. Vyukov, ``AddressSanitizer: A Fast Address Sanity Checker,'' USENIX ATC, 2012.
\bibitem{klee} C. Cadar, D. Dunbar, and D. Engler, ``KLEE: Unassisted and Automatic Generation of High-Coverage Tests for Complex Systems Programs,'' OSDI, 2008.
\bibitem{dullien-weird-machines} T. Dullien, ``Weird Machines, Exploitability, and Provable Unexploitability,'' \emph{IEEE Transactions on Emerging Topics in Computing}, vol. 8, no. 2, pp. 391--403, 2020, doi: 10.1109/TETC.2017.2785299.
\end{thebibliography}
\end{document}